\PassOptionsToPackage{utf8}{inputenc}
\documentclass{bioinfo}
\copyrightyear{2015} \pubyear{2015}

\access{Advance Access Publication Date: Day Month Year}
\appnotes{Manuscript Category}

\def\bSig\mathbf{\Sigma}

\usepackage[figuresright]{rotating}
\usepackage{amsmath}
\usepackage{algorithm}  
\usepackage{algorithmic} 

\usepackage{bm}

\usepackage{graphicx}
\usepackage{float}
 
\usepackage{subfigure}

\usepackage{multirow}   
\usepackage{threeparttable}  
\usepackage{makecell}

\usepackage{soul} 
\usepackage{color, xcolor}
\soulregister{\cite}7  
\soulregister{\citep}7  
\soulregister{\citet}7  
\soulregister{\ref}7  
\soulregister{\pageref}7  
\soulregister{\url}7

\begin{document}
\firstpage{1}

\subtitle{Subject Section}

\title[ ]{STW-MD: A Novel Spatio-Temporal Weighting and Multi-Step Decision Tree Method for Considering Spatial Heterogeneity in Brain Gene Expression Data}

\author[Sample \textit{et~al}.]{Shanjun Mao\,$^{\text{\sfb 1}}$, 
Xiao Huang\,$^{\text{\sfb 1}}$, 
Chenyang Zhang\,$^{\text{\sfb 1}}$, 
Runjiu Chen\,$^{\text{\sfb 1}}$, 
Yizhu Diao\,$^{\text{\sfb 1}}$, 
Zongjin Li\,$^{\text{\sfb 2}}$, 
Qingzhe Wang\,$^{\text{\sfb 3}}$,
Shan Tang\,$^{\text{\sfb 1,}*}$ 
and Shuixia Guo\,$^{\text{\sfb 4, \sfb 5}*}$}
\address{$^{\text{\sf 1}}$College of Finance and Statistics, Hunan University, Changsha 410079, China \\
    $^{\text{\sf 2}}$School of Statistics and Mathematics, Central University of Finance and Economics, Beijing 100081, China\\
    $^{\text{\sf 3}}$College of Finance and Statistics, Hunan University, Changsha 410079, China\\
    $^{\text{\sf 4}}$MOE-LCSM, School of Mathematics and Statistics, Hunan Normal University, Changsha 410081, China \\
    $^{\text{\sf 5}}$Key Laboratory of Applied Statistics and Data Science, Hunan Normal University, College of Hunan Province, Changsha 410081, China}

\corresp{$^\ast$Corresponding authors. College of Finance and Statistics, Hunan University, Shijiachong Road, Changsha 410000, China. E-mail: tangshanhaha@hnu.edu.cn (S.T.); MOE-LCSM, School of Mathematics and Statistics, Hunan Normal University, Lushan Road, Changsha 410000, China. E-mail: guoshuixia75@163.com (S.G.).}

\history{Received on XXXXX; revised on XXXXX; accepted on XXXXX}

\editor{Associate Editor: XXXXXXX}

\abstract{
    \textbf{Motivation:} 
    Gene expression during brain development or abnormal development is a biological process that is highly dynamic in spatio and temporal. Due to the lack of comprehensive integration of spatial and temporal dimensions of brain gene expression data, previous studies have mainly focused on individual brain regions or a certain developmental stage. Our motivation is to address this gap by incorporating spatio-temporal information to gain a more complete understanding of the mechanisms underlying brain development or disorders associated with abnormal brain development, such as Alzheimer's disease (AD), and to identify potential determinants of response.\\
    \textbf{Results:}
    In this study, we propose a novel two-step framework based on spatial-temporal information weighting and multi-step decision trees, which can effectively exploit the spatial similarity and temporal dependence between different stages and different brain regions, and facilitate differential gene analysis in brain regions with high heterogeneity problems. In this paper, we focus on two datasets: the AD dataset includes gene expression data from early, middle, and late stages, while the brain development dataset spans fetal development to adulthood. The results show that the model framework in this paper can effectively analyze the genes affecting brain development and AD progression in different brain regions and different stages, which is consistent with some existing studies, and can provide some insights into the process of brain development and abnormal brain development behavior.\\
    \textbf{Availability:} STW-MD is available at. \\
    \textbf{Contact:} \href{name@bio.com}{name@bio.com}\\
    \textbf{Supplementary information:} Supplementary data are available at \textit{Bioinformatics}
    online.}

\maketitle

\section{Introduction}

Human brain development is a dynamic and highly regulated biological process that unfolds over a protracted period.   \citep{li2018integrative, de2018dynamic,boyce2020genes}. Brain development and function depend on the precise regulation of gene expression, and the regulatory processes of different genes are constantly changing from fetal development to adulthood \citep{weyn2018precise}. With a focus on brain development in adolescence, \citet{gao2019review} had investigated the gene regulation processes responsible for the significant maturation of brain development from childhood to adulthood. Studying gene expression trajectories during brain development can help us understand many advanced behavioral functions in humans, such as the abilities of learning and memory \citep{li2020evolution,tang2020gene}. Meanwhile, abnormal brain development at different stages can also lead to serious diseases, such as Autism spectrum disorder (ASD) with early childhood onset \citep{bicker2021criss}, schizophrenia with late adolescence or early adulthood \citep{ma2019multi}, and Alzheimer's disease (AD) in older age groups \citep{zhang2022clinical}. These disorders of brain development, with a high medical burden, have attracted global attention. Among them, AD is a progressive and irreversible neurodegenerative disease that affects cognition, function, and behavior. The progress of AD is a continuous process from preclinical disease, mild cognitive and/or behavior disorder, dementia and alzheimer's disease \citep{porsteinsson2021diagnosis}. 
Therefore, the study of differential gene expression in different periods of brain development or different stages of related diseases helps to promote the understanding of brain development and disease mechanism.

In essence, the gene expression data related to brain development or abnormal development can be summarized as a kind of spatial-temporal dynamic evolution data, that is, there are multiple different periods or stages, and the gene expression data of different brain regions. With the rapid development of modern biotechnology, richer and more detailed spatio-temporal data of the brain have been obtained. For example, \citet{jiao2019brainexp} proposes a Database called Brain EXPression Database (BrainEXP) that provides basic gene expression across specific brain regions, age ranges, and genders. In different periods and regions of brain development, different gene expression levels may have different regulatory effects. \citet{lopes2022genetic} stated that the transcriptional heterogeneity of human microglia varies by brain region and aging. Furthermore, insights from AD highlight the importance of investigating regional alterations in brain structure and functionality for elucidating disease pathogenesis. \citet{Blair2020cytoarchitectonic} found that gray matter density (GMD) in the hippocampus, amygdala, and basal forebrain decreased faster in AD patients than in healthy control subjects, and subcortical regions also decreased faster than neocortical brain regions. Understanding gene regulation across multiple brain regions and developmental stages is therefore crucial to provide mechanistic insights into brain development or abnormal development.

The spatio-temporal data of brain development mainly involve two questions: One is to study which genes play a role in brain development or developmental abnormalities. The other is to study which brain regions and during what developmental time periods these genes play a role. Various approaches have been developed to address the temporal and spatial dynamics data of brain development and/or abnormal development. 
\citet{lin2015markov} proposed a two-step approach based on Markov random fields (MRF) to effectively utilize the spatio-temporal information embedded in brain regions and improve the ability to identify differentially expressed (DE) genes during brain development. \citet{semick2019integrated} investigated paired DNAm and transcriptome data from four brain regions in AD patients, and linked methylation differences to local gene dysregulation through brain region stratification analysis and cross-region differential methylation analysis. \citet{jung2021novel} constructed a transcriptome-based weighting polygenic risk score (TW-PRS) for each brain region and the MultiXcan statistical method was used to integrate the results of transcriptome wide association studies (TWAS) in these 13 brain regions, which provided additional information for identifying individuals at high risk for AD \citep{barbeira2019integrating}. However, the above methods either only consider the brain region factor without considering the time scale factor, or they consider both but ignore the gene heterogeneity during brain development, that is, the expression level of the same gene is not consistent in different regions of the brain.
Regardless of the clinical presentation, both neurological and psychiatric disorders demonstrate high individual and regional heterogeneity in gene expression patterns \citep{hampel2023foundation}. For example, in AD patients, gene expression patterns in microglia exhibit high dynamics and heterogeneity due to epigenetic modifications and non-coding RNA regulation \citep{li2022heterogeneity}. Therefore, it is necessary to address the above issues and consider the heterogeneity of genes when studying the spatio-temporal dynamics data of brain development.

Previous studies of gene expression data for different stages of brain development or brain diseases have either considered only the temporal nature of the data and different developmental stages, or only the spatial dimension of different brain regions. Concomitantly, owing to the inherent heterogeneity within gene expression data, namely the substantial diversities in functional contributions of genes in distinct brain regions over the course of development or disease progression, a direct differential expression analysis may not adequately account for these sources of variability. Considering the spatio-temporal dynamics of gene expression data during brain development or abnormal development, a two-step modeling framework based on spatio-temporal information weighting and multi-step decision trees (STW-MD) is proposed in this paper. Specifically, the framework first weighted the gene expression data of different brain regions based on the differentially expressed gene information at different stages to cover the differences between different brain regions and different periods. Then, a multi-step decision tree model was used to screen key gene modules, and their biological significance was interpreted through gene enrichment analysis. Concurrently, as an efficient and interpretable method, the framework can well adapt to some existing differential gene analysis methods and clustering algorithms. The methodology is described in Section 2. Section 3 demonstrates the application of the proposed framework to Alzheimer's disease datasets and neurodevelopmental data, presenting key insights generated. Finally, Section 4 concludes the paper.

\section{Materials and methods}
\label{s:model}

\subsection{Datasets}

To demonstrate the effectiveness of the proposed method under the spatio-temporal dynamic data of the brain, this study intends to conduct research based on two datasets: the AD dataset and the brain development dataset. The former was obtained from the Mount Sinai Medical Center Brain Bank (MSBB), which included data from 118 samples (27 healthy controls and 91 AD patient samples) \citep{guo2023sex,wang2016integrative}. 
Each AD sample contained data on the expression levels of 18534 genes in 19 brain regions from early, middle to late stages. The staging criteria primarily relied on the Clinical Dementia Rating (CDR) Scale. The brain development dataset was obtained from the BrainSpan database \citep{ma2019multi,lin2015markov,kang2011spatio}, which was collected from 1340 tissue samples of 57 developmental and postmortem adult brains and contains gene expression data of 16 brain regions at 15 stages of brain development from embryonic development to late adulthood. After excluding non-coding genes and lowly expressed genes, a total of 15210 genes were retained as the background for further analysis. Details and features of the two datasets are provided in Supplementary Material S1.

Suppose different periods or stages of brain development are defined as $(0, 1,2,\cdots,T, T+1,\cdots)$. Let $y_{t,i,r,g}$ denote the observed gene expression value for gene $g$ in the $i$-th samples in brain region ${r}$ and period $t$, and let $\mathbf{N}=(N_{0},N_{1},\ldots,N_{T},N_{T+1},\ldots)$ denote the number of samples in different periods.

\subsection{Methods}



\begin{figure*}[h]
    \centering
    \includegraphics[width=15cm]{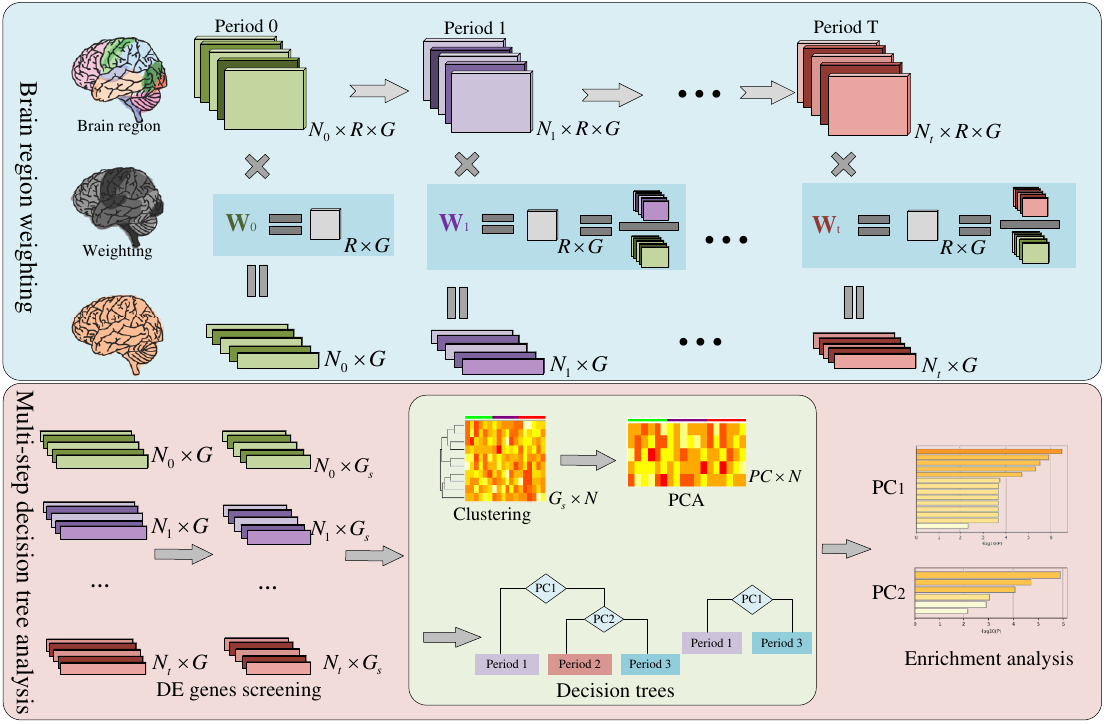}
    \caption{Temporal and spatial dynamic evolution framework of gene expression in the brain. 
    A Framework for Investigating Spatio-Temporal Dynamics of Gene Expression during Brain Development. 
    The proposed model framework is categorized into two distinct components: brain region weighting (top panel, light blue area) and multi-step decision tree analysis (bottom panel, light red area). The initial layer of the brain region weighting section encompasses brain gene expression samples derived from various evolutionary periods across brain development. Each distinct sample is visually represented as an $R \times G$ matrix block, where the rows and columns correspond to diverse brain regions and genes, respectively. Subsequently, the second layer involves the calculation of weights assigned to distinct brain regions. These weights are primarily derived from the comparative analysis of sample information between the current period's brain regions and the period $0$. The third layer illustrates the results of weighted gene expression, with each block denoting the gene expression of an individual sample. 
    The multi-step decision tree analysis was divided into three main parts: the DE gene screening process, the clustering and decision tree process (light green area) and the enrichment analysis process.}
    \label{fig02}
\end{figure*}

\subsubsection{Weighting Brain Regions by Considering Genetic Heterogeneity}

This paper proposes a novel weighting method for spatio-temporal dynamic expression data, which constructs weights based on the differential expression information of different stages, and then weights all brain regions in the same stage, that is, reduces the dimension of all brain regions in spatial sense, so that we can combine the data of all brain regions to study gene expression at different stages of the brain. The method of weighting spatio-temporal information is illustrated graphically, as shown in the light blue part of Figure~\ref{fig02}.

Conventional weighting approaches employ dimensionality reduction techniques similar to arithmetic averaging to weight the spatial-temporal gene expression data \citep{li2021improved,crowell2020muscat}. However, such strategies result in substantial loss of the original information, and fail to adequately address the problem of gene heterogeneity in different brain regions. Therefore, the Fold Change (FC) was utilized as the weighting factor to integrate gene expression profiles across different brain regions in every stage \citep{dembele2014fold,mandelboum2019recurrent}. FC was principally computed from the ratio of average gene abundances between two different stages, and each gene was corresponding to different weights in different stages and different brain regions. By selecting FC as the weighting metric, inter-genic, inter-regional and temporally divergent variability could be holistically encompassed. This strategically amalgamated spatial and temporal dimensions for comprehensive analysis of brain development dataset.

Therefore, the weight FC value of brain region $r$ in period $t$ is as follows:

\begin{equation}
    FC_{t, r, g}=\begin{cases}
        \frac{\sum_{i_t}^{N_t} y_{t,i_t,r,g}/N_t }{\sum_{i_{0}}^{N_{0}} y_{0,i_{0},r,g}/N_{0} } & \text{if}\; t \geq 1\\
        1 & \text{if}\; t = 0 
    \end{cases}
    \label{eq:01}
\end{equation}
where $y_{t,i_t,r,g}$ and $y_{0,i_{0},r,g}$ are the gene expression values, and $N_t$ and $N_{0}$ represent the sample size in the corresponding period.
Meanwhile, period $t=0$ can be considered as the control group of period $t \geq 1$, for example when studying gene expression analysis in the onset stage of AD, period $t=0$ represents the gene expression level in healthy individuals. 

FC value is usually used to identify differentially expressed genes (DEGs) between two diseases, but $\log_2 FC$ is more suitable to deal with asymmetric expression differences than FC value, and can provide a more accurate measure of genes with low expression. Here, we use the $\log_2 FC$ value as the weight of brain regions, which can well take into account the differences in developmental evolution of different genes in different brain regions. Hence, the brain region gene expression of sample $i_{t}$ in period $t$ was weighted as follows:
\begin{equation}
    \begin{aligned}
        y_{t,i_t,g}^{FC} = \frac{\sum_{r_t}^{R} y_{t,i_t,r_t,g} \times \log_2 FC_{t, r_t, g} }{\sum_{r_t}^{R} \log_2 FC_{t, r_t, g}}
    \end{aligned}
    \label{eq:02}
\end{equation}
where $y_{t,i_t,g}^{FC}$ represents the weighted gene expression of gene $g$ corresponding to individual $i_t$ in period $t$ over $R$ brain regions. The weighted gene expression data is the result after taking into account gene heterogeneity and gene expression differences in different brain regions, which can be subsequently analyzed using some existing gene expression analysis.

At the same time, considering that the same gene may be up-regulated and down-regulated simultaneously in different brain regions, direct weighting will cause a situation where positive and negative cancel each other and "neutralize" differential gene expression. Therefore, the weights defined above need to be further adjusted to make the adjusted weights better distinguish the gene expression levels in different periods after weighting. However, because the status of up-regulated and down-regulated genes is uncertain, the way of weight adjustment also varies due to the above situation. A detailed discussion is provided in Supplementary Material S2.1.
And because the threshold value of FC is usually set to 1.5 in commonly used differential gene screening, the threshold selection of weights in this paper is also set to 1.5. In this case, the adjusted weights are shown as follows:
\begin{equation} 
    \log_2 FC_{g, r, n}^{ad} \triangleq \begin{cases}
        -\log_2 FC_{g, r, n} & \text { if } \log_2 FC_{g, r, n}< - \log_2 1.5 \\  
        \log_2 FC_{g, r, n} & \text { if } \log_2 FC_{g, r, n} \geq \log_2 1.5 \\
        0 & \text {others.}
    \end{cases}   
    \label{eq:03}
\end{equation}
where $\log_2 FC_{g, r, n}^{ad}$ represents the adjusted weight, $\log_2 F C_{g, r, n} \geq \log_2 1.5$ indicates up-regulated genes, $\log_2 F C_{g, r, n}< - \log_2 1.5$ indicates down-regulated genes, and the others are considered insignificant.
For the case that the gene is not significant in the brain region, if its weight is not set to 0, the significant expression level of the gene will be "diluted" by the insignificant differentially expressed gene levels in other brain regions during the weighting process, thus causing an unreasonable situation. 
More details are discussed can be found in Supplementary Material S2.2.

In addition, as the expression values of the same gene differ greatly in different brain regions. To reduce the impact of this phenomenon on brain region weighting in subsequent analysis, the gene expression values in the matrix were first standardized. Specifically, the expression values of the same gene in the same sample from different brain regions were normalized, and the expression values were controlled within the range of [1,10]. Then the normalized expression level of gene $g$ at period $t$ is as follows:
\begin{equation}
    \begin{aligned}
        y_{t,i,r,g}^{\prime}=\frac{y_{t,i,r,g}-\underset{r \in (1,2,\dots,R)}{\min}  y_{t,i,r,g}}{\underset{r \in (1,2,\dots,R)}{\max}  y_{t,i,r,g} -\underset{r \in (1,2,\dots,R)}{\min} y_{t,i,r,g}} \times 3 + 1
    \end{aligned}
    \label{eq:04}
\end{equation}
where $y_{t,i,r,g}^{\prime}$ represents the normalized result, and $\underset{r \in (1,2,\dots,R)}{\max} y_{t,i,r,g}$ and $\underset{r \in (1,2,\dots,R)}{\min} y_{t,i,r,g} $ represent the maximum and minimum gene expression levels in $R$ brain regions of period $t$, respectively.

To concisely recapitulate the sequential steps comprising the present methodology, they may be delineated as follows:
\begin{itemize}
    \item Step 1: Normalize $\bm{y}$ to $\bm{y}^{\prime}$ by Formula~\eqref{eq:04};
    \item Step 2: Calculated adjusted FC value by Formula~\eqref{eq:03};
    \item Step 3: Calculate weighted gene expression by Formula~\eqref{eq:02}.
\end{itemize} 

\subsubsection{Multi-step Decision Tree based on gene clusters}

In the previous section, gene expression data of individuals at different periods were obtained by weighting the data of brain regions. In the next step, some common differential gene analysis methods, such as t-test, SAM method, Limma model and GO/KEGG enrichment analysis \citep{dai2020integrated,yu2023identification}, can be combined to identify the key genes that distinguish different periods of brain development or disease progression. In this paper, a strategy of integrating multiple methods was used for analysis, that is, a multi-step decision tree method was used for differential gene expression analysis. The algorithm for the multi-step decision tree is as follows (see pink part of Figure~\ref{fig02}):

\begin{itemize}
    \item Screening differentially expressed genes. In this paper, differentially expressed genes were selected from gene expression data mainly based on the commonly used Limma model \citep{silva2023discovery,ritchie2015limma}. The first dimension reduction can be achieved by selecting differentially expressed genes.
    \item Cluster analysis and matrix factorization. The differentially expressed genes screened above were grouped into several gene modules by cluster analysis, such as tSNE or UMAP combined with K-means clustering methods \citep{jorgensen2020age,hozumi2021umap}. Combined with principal component analysis (PCA) \citep{lopez2011principal}, the principal components of each gene class were extracted to achieve the quadratic dimension reduction, and the interpretability of the subsequent decision tree model was enhanced.
    \item The decision tree model. The PC of the above gene classes were used in the CART decision tree model to identify the key gene classes that distinguished AD patients at different stages \citep{costa2022decision}.
    \item Gene Set Enrichment Analysis (GSEA). GO or KEGG enrichment analysis \citep{su2022identification,kramarz2019gene} was used to analyze the relationship between the above key gene categories and brain development or disease progression.
\end{itemize}

\section{Results}
\label{s:result}

\subsection{Exploring the Influential Factors of Alzheimer's Disease}

\subsubsection{Weighting results for significantly different genes based on FC values}

\begin{figure}[h]
    \centering
    \includegraphics[width=7.5cm]{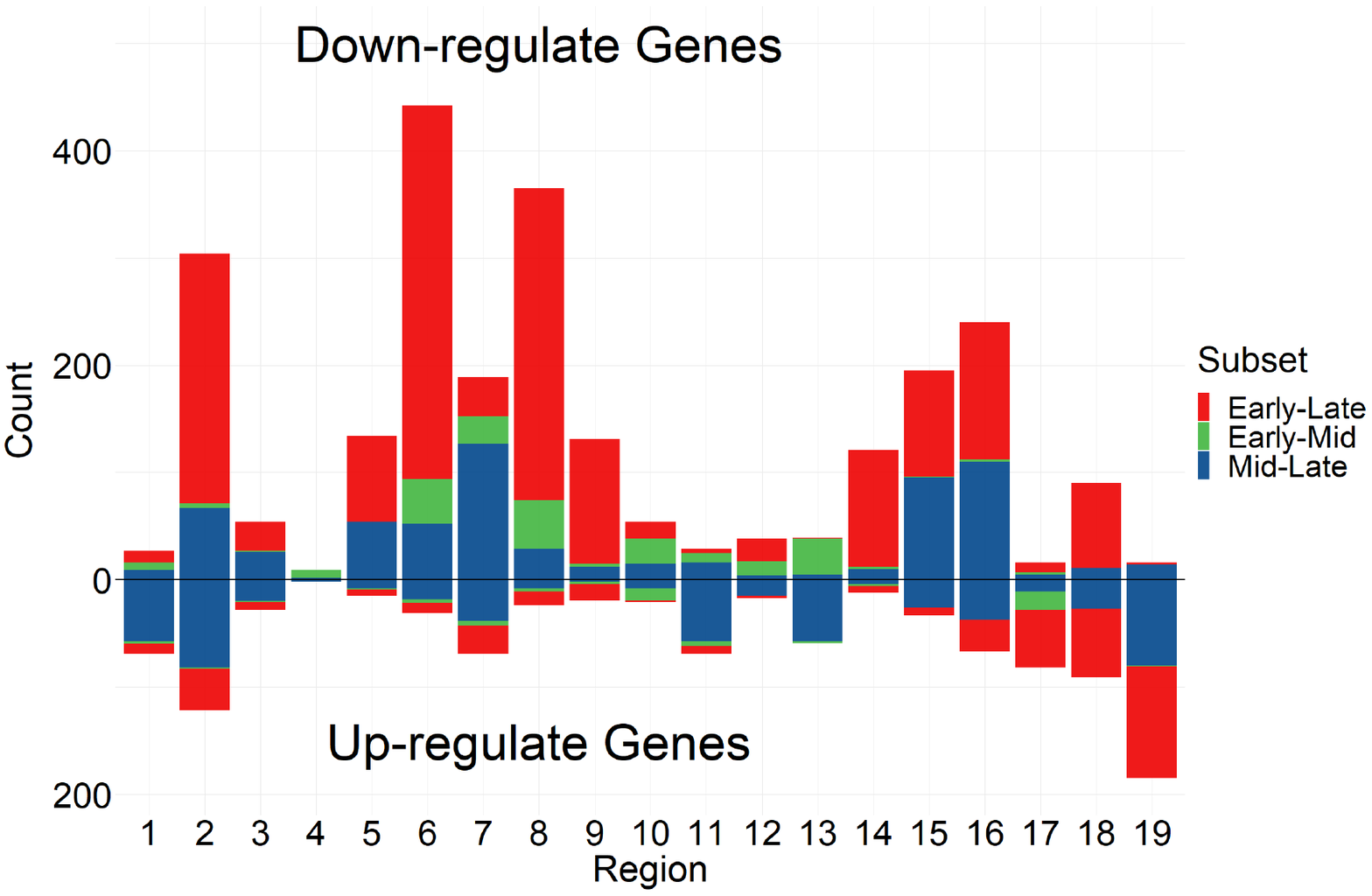}
    \caption{Bar graph of level grouping of DEGs in 19 brain regions. The bars in the upper and lower halves represent the number of down-regulated and up-regulated genes, respectively. The different colors indicate the DEGs between pairs at various stages of AD.}
    \label{fig10}
\end{figure}

In AD research data, the expression levels of most genes may not be significantly different in different stages, so it is not worthwhile to calculate the weight for these genes. In this paper, Limma model was used to conduct a preliminary differential gene screen for gene expression levels in different time stages in each brain region. Figure~\ref{fig10} displays bar graphs depicting the number of up-regulated and down-regulated genes in different brain regions of the AD dataset. Each column is color-coded to represent the count of differentially expressed genes between pairs of the three stages. Among them, the differential expression of genes was predominantly observed in the early-late and middle-late groups. After pooling the differentially expressed genes in each brain region, the number of significantly up-regulated genes was significantly less than that of significantly down-regulated genes (a total of 458 up-regulated genes and 1830 down-regulated genes), which was the main reason for the adjustment of weights in the above methods. Taking the union of all DEGs from the 19 brain regions, we obtained a total of 1014 DEGs. 

\begin{figure}[h]
    \centering
    \subfigure[Before standardizations]{\includegraphics[width=0.23\textwidth]{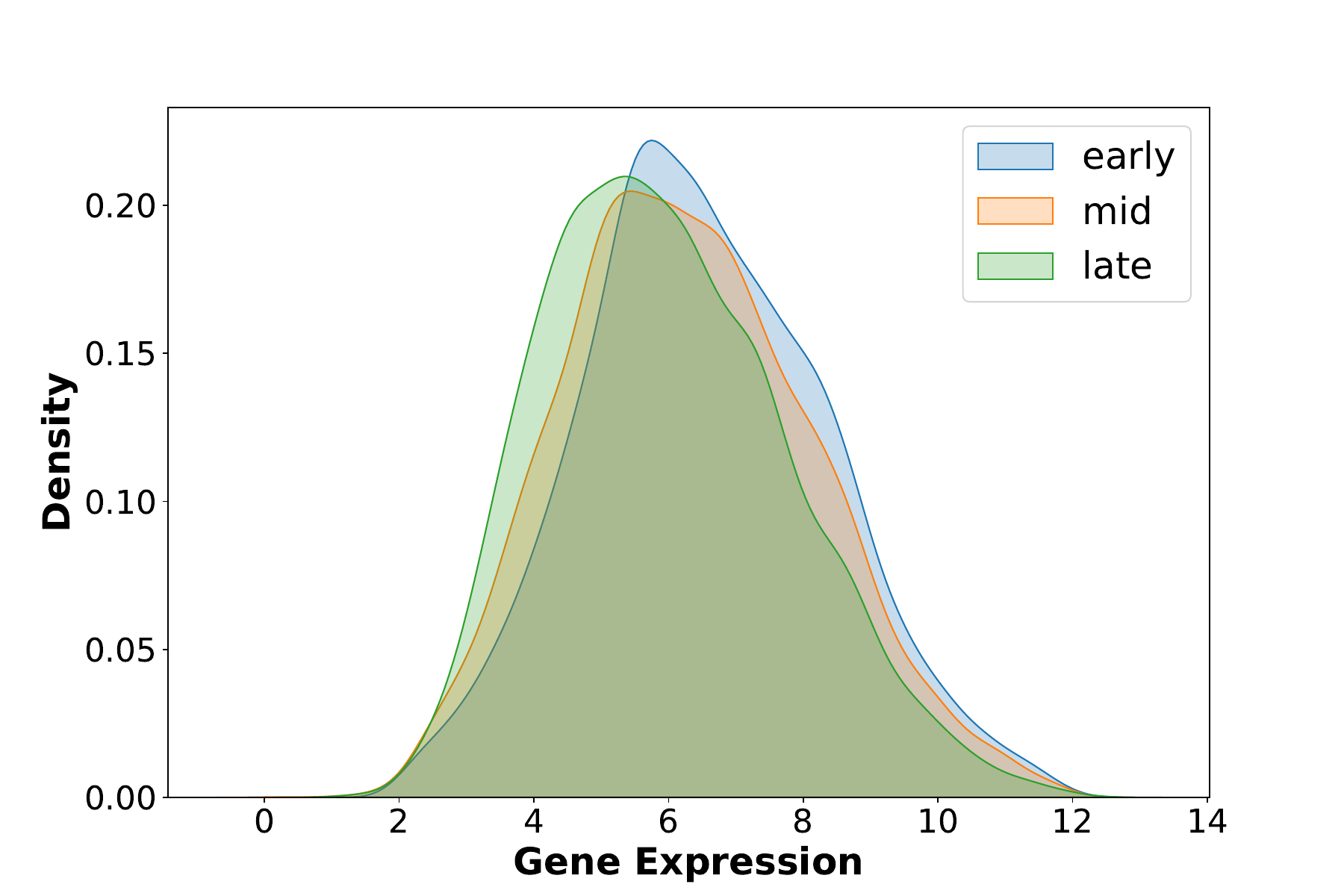}}
    \subfigure[After standardization]{\includegraphics[width=0.23\textwidth]{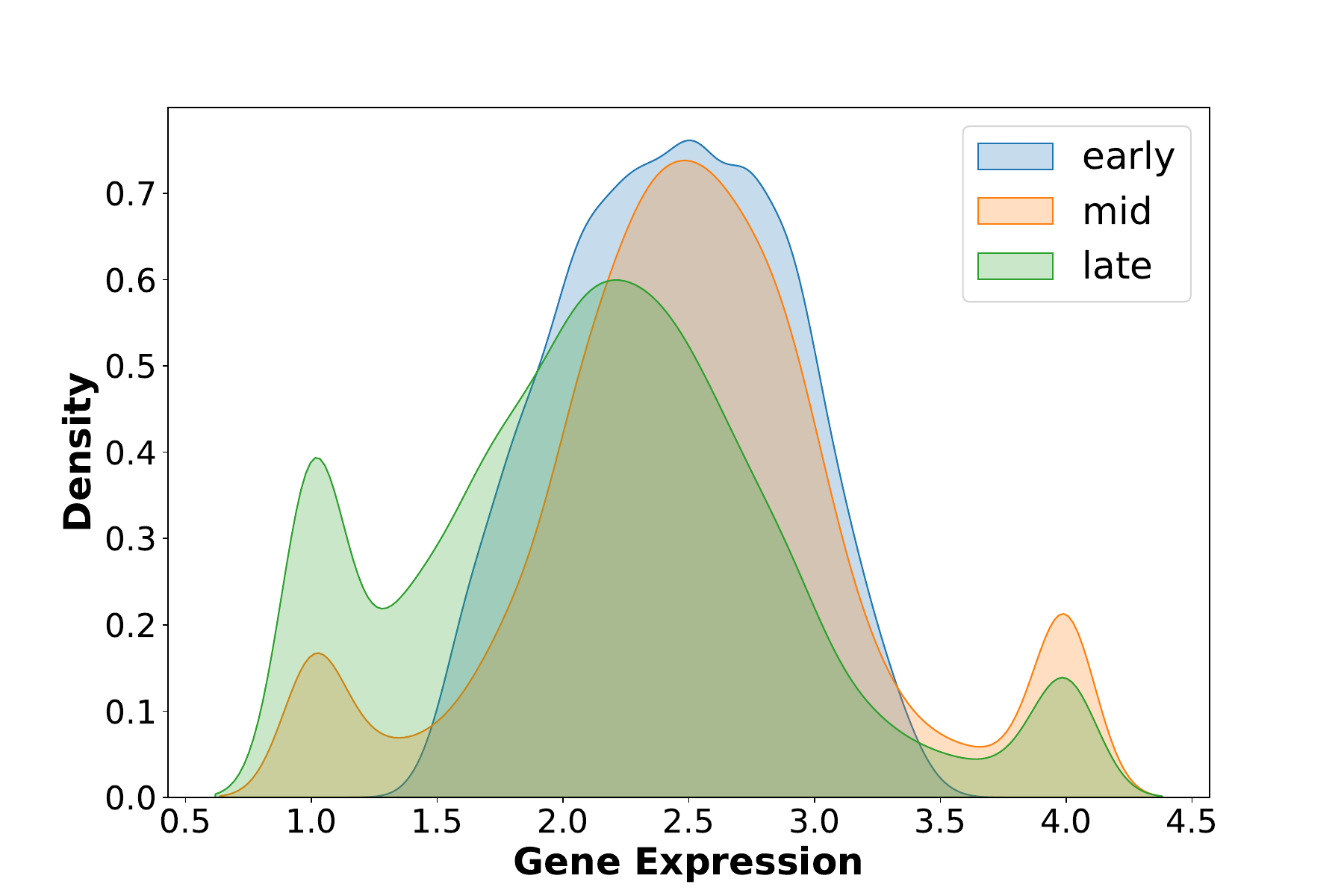}}
    \caption{Comparison density map of normalization treatment or not.}
    \label{fig05}
\end{figure}

Based on the adjusted FC value, the differentially expressed genes in 19 brain regions of AD patients were weighted after standardization. Figure ~\ref{fig05} shows the results before and after weighting and standardized distribution, it can be seen in different stages in patients with AD become more significant difference between groups, is advantageous to the genetic difference of subsequent analysis. Details are discussed and illustrated in Supplementary Material S2.1.

\subsubsection{Differential gene expression analysis results after Weighting}

Differential gene expression analysis was performed on the weighted gene expression data. 
A total of 50 differentially expressed genes (23 up-regulated and 27 down-regulated genes) were identified in the early and middle stages, 87 differentially expressed genes (14 up-regulated and 73 down-regulated genes) in the early and late stages, and 110 differentially expressed genes (16 up-regulated and 94 down-regulated genes) in the middle and late stages. Overall, the differentially expressed genes significantly lower. The combination of the three groups of differentially expressed genes in 158 genes. These three groups of differentially expressed genes were combined, and a total of 158 differentially expressed genes were obtained.

\begin{figure*}[h]
    \centering
    \subfigure[The clustering results]{\includegraphics[width=0.3\textwidth]{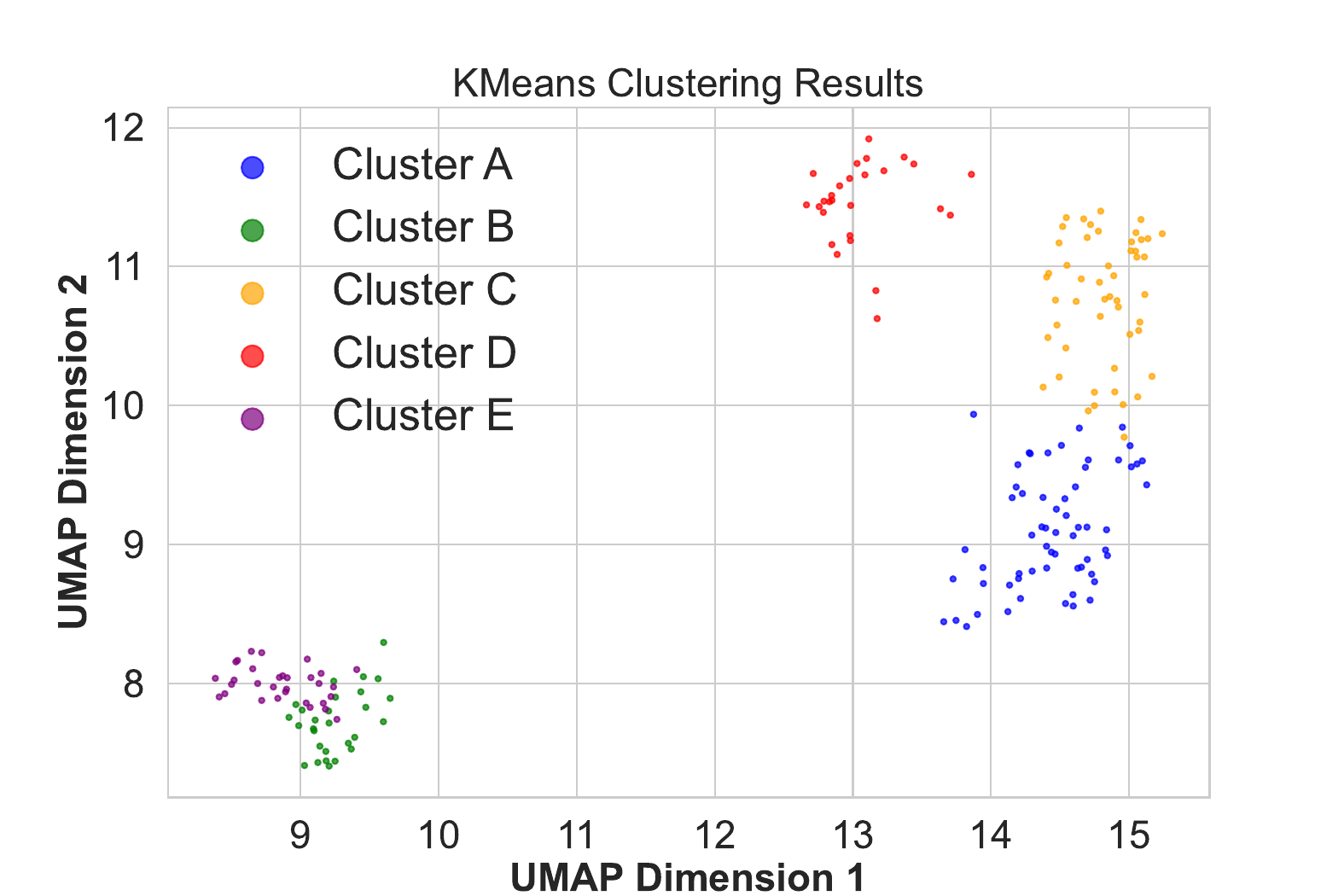}}
    \subfigure[The number of DEGs]{\includegraphics[width=0.3\textwidth]{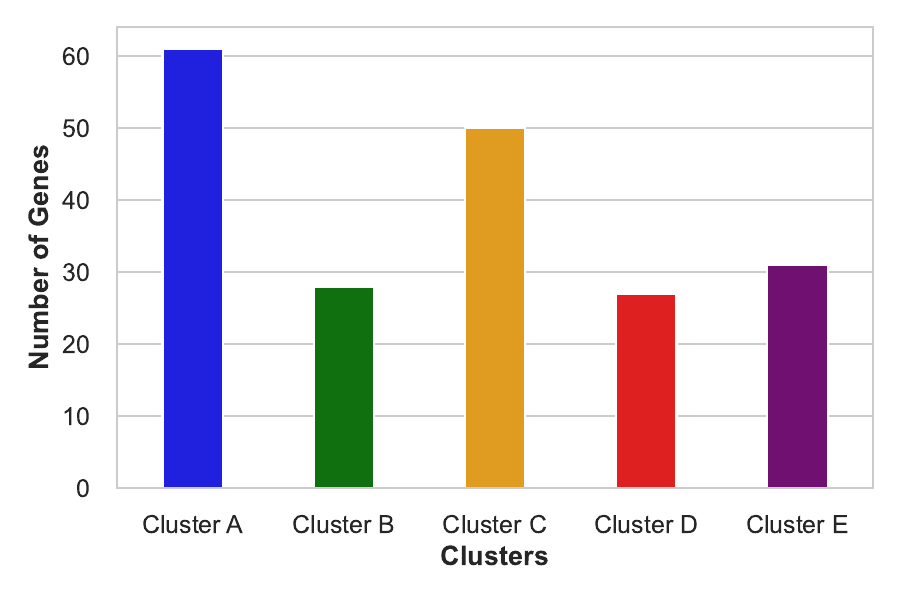}}
    \subfigure[The VCRate after PCA]{\includegraphics[width=0.3\textwidth]{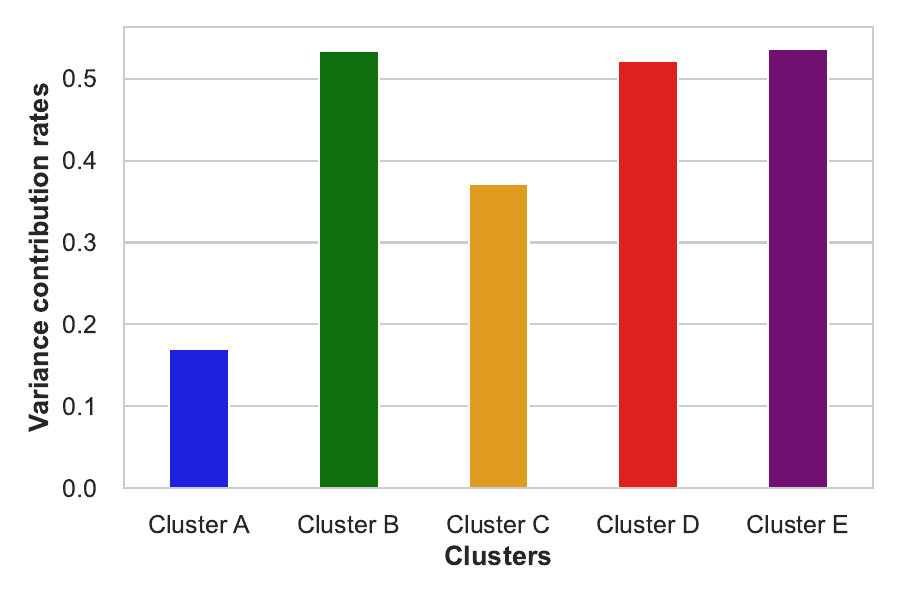}}

    \caption{
    The result of the multi-step decision tree algorithm. The first two panels are the clustering results and the number of differentially expressed genes in each category, and the last panel is the variance contribution rate of the first principal component after PCA for each cluster gene.}
    \label{fig055}
\end{figure*}

The first two panels in the second row of Figure~\ref{fig055} present the results of the cluster analysis of differentially expressed genes, where it is appropriate to divide them into five classes based on certain criteria of kmeans and defined as ClusterA, ClusterB, ClusterC, ClusterD, and ClusterE, respectively. The last panel shows the variance contribution rate of the principal component of the first principal component of the five groups of differentially expressed genes, among which the variance contribution rate of the first principal component in each group of the five groups was mostly above 50\%, indicating that the subsequent analysis based on the first principal component was feasible.

\begin{figure*}[h]
    \centering
    \includegraphics[width=14cm]{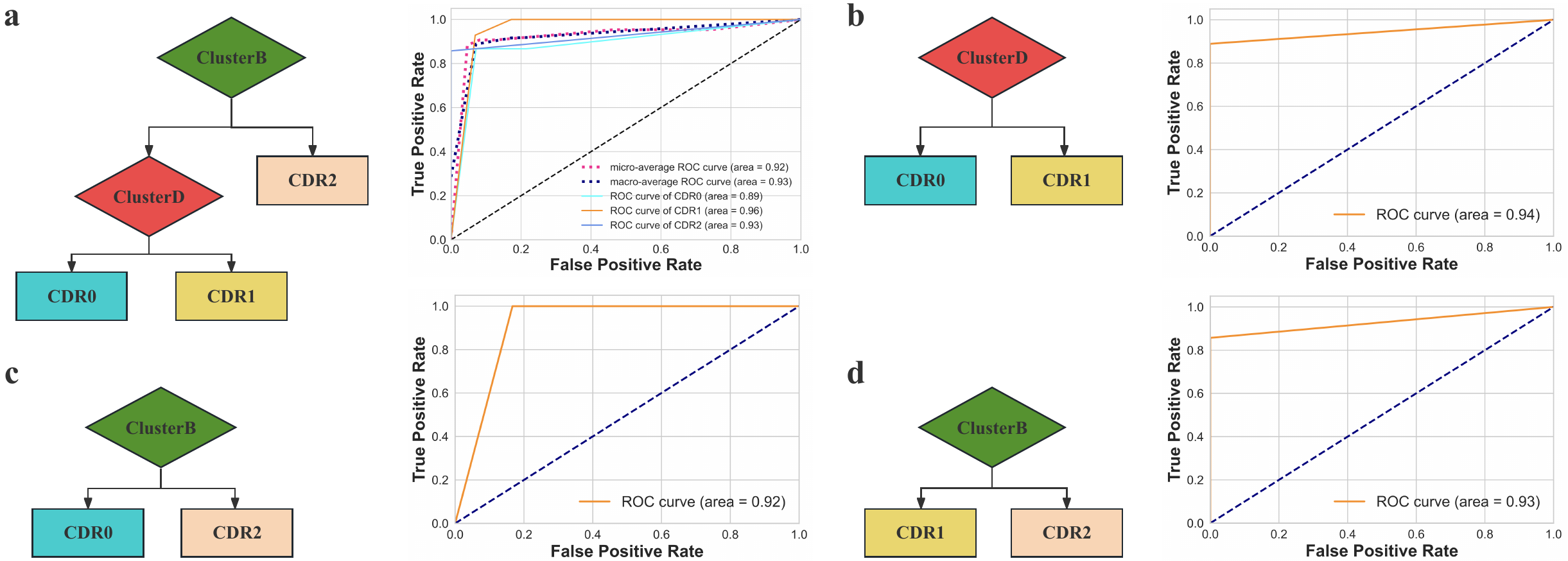}
    \caption{Early, mid, and late decision trees and their classification ROC curves. Among them, diamonds represent taxonomic features (ClusterB and ClusterD), and rectangular blocks (Early, mid and late) represent AD individuals in different stages.}
    \label{fig14}
\end{figure*}

Based on the first principal component data of the above five groups of differential genes, Figure~\ref{fig14} shows the classification results of different stages using the decision tree model \citep{saint2023disease}. First of all, the first panel of Figure~\ref{fig14} shows the decision tree of the early, middle and late stages, where the selection complexity coefficient CP=0.05 is used to trim the decision tree. The results show that the final classification accuracy of 20 times five-fold cross validation is 88.73\%, indicating that the model can well distinguish different stages of AD patients. In addition, the pairwise decision trees of early, middle and late stage patients were constructed, and the prediction accuracy reached 90.41\%, 87.94\% and 92.42\%, respectively, as shown in the last three panels of Figure~\ref{fig14}, in which the principal components of ClusterB and ClusterD differential genomes played an important role in the classification of AD patients. It can be seen that the principal component of the ClusterB differential gene class is an important feature to distinguish early or middle stage AD patients from late stage AD patients, and the principal component of the ClusterD differential gene class is an important feature to distinguish patients with early and middle AD. 

Meanwhile, Table~\ref{tab:4} provides a comparative evaluation of the proposed spatio-temporal brain region weighting algorithm against two additional approaches - namely brain region averaging and brain region imputation. The results demonstrate that the classification performance attained with the developed algorithm was notably superior to those achieved by the alternative methods, as evidenced by improved accuracy metrics. mcie package.

\begin{table}[!htbp]
    \centering
    \vspace*{-6pt}
    \caption{Comparison Results.}
    \begin{tabular}{@{}lccc@{}}
    \hline
    Comparison     & Averaging  & Imputation  & Weighted  \\ \hline
    early   VS mid VS late & 49.41\%        & 47.09\%    & \bm{$88.73\%$}                  \\
    early   VS mid         & 70.55\%        & 52.91\%    & \bm{$90.41\%$}                 \\
    early   VS late        & 64.24\%        & 51.06\%    & \bm{$87.94\%$}                  \\
    mid   VS late          & 63.33\%        & 61.67\%    & \bm{$92.42\%$}                  \\ \hline
    \end{tabular}
    \label{tab:4}%
    \vskip12pt
\end{table}

\subsubsection{Enrichment analysis results of ClusterB and ClusterD genes}

\begin{figure*}[h]
    \centering
    \subfigure{\includegraphics[width=0.48\textwidth]{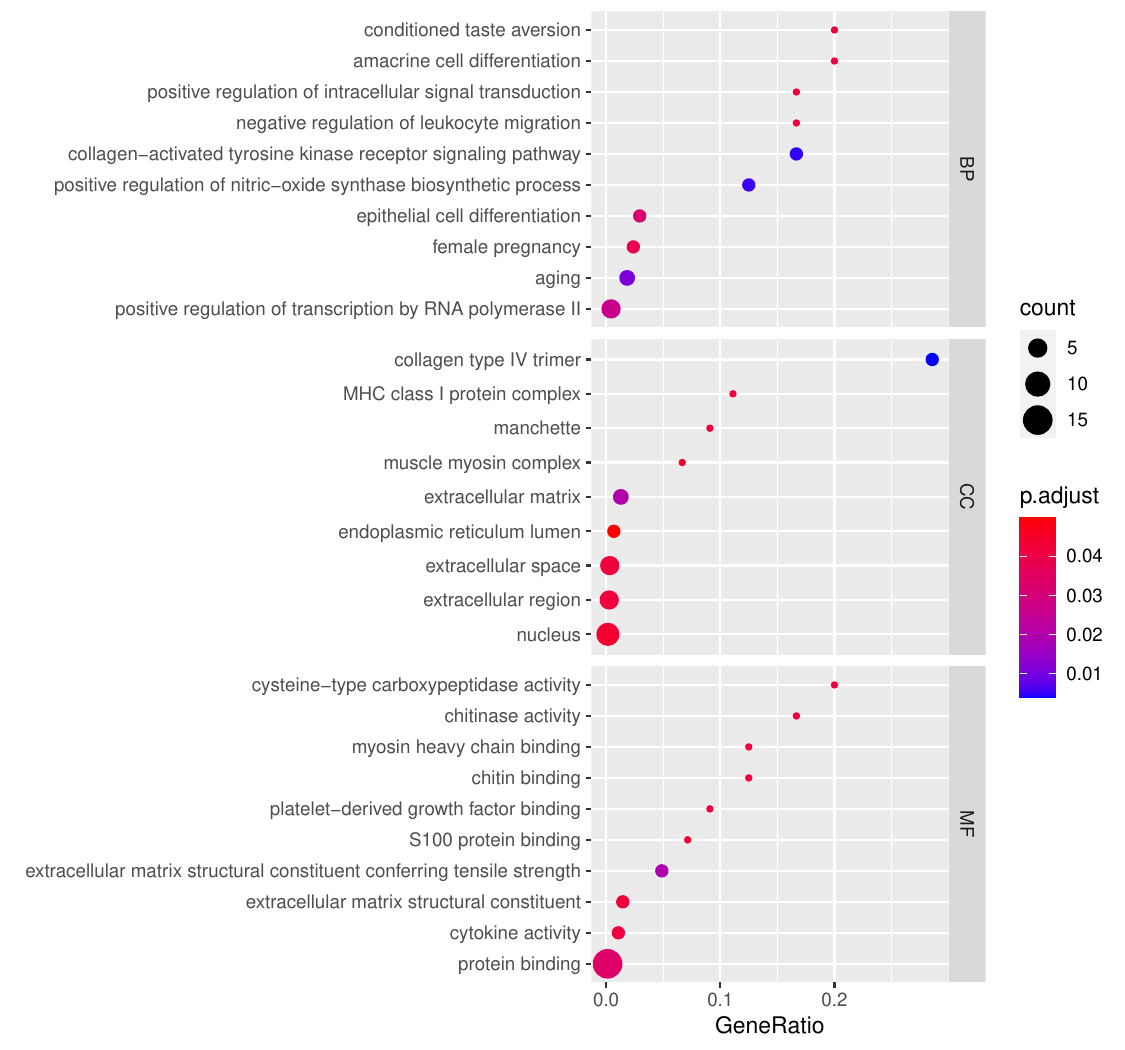}}
    \subfigure{\includegraphics[width=0.48\textwidth]{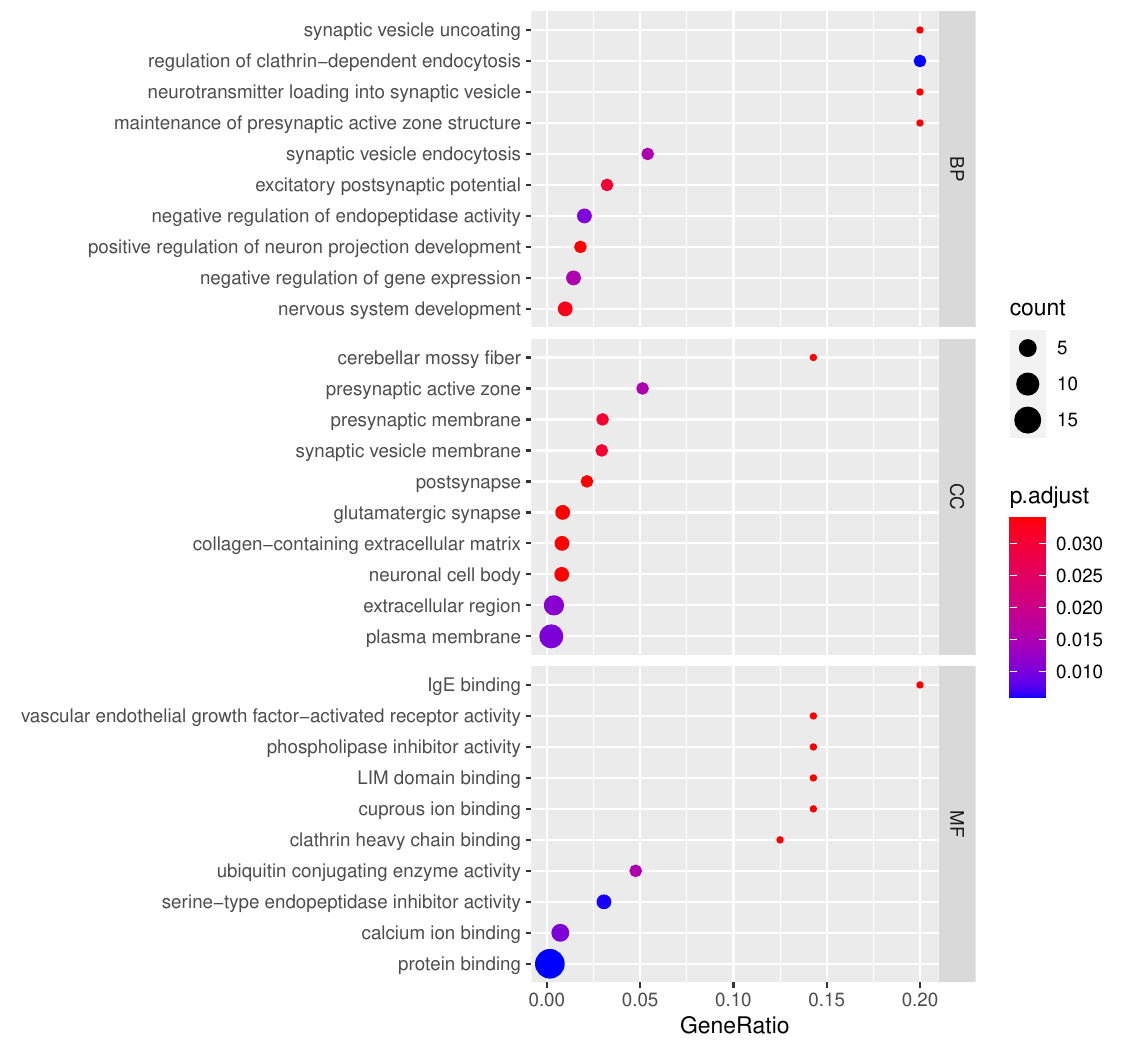}}
    \caption{Bubble graphs depicting the top 10 most enriched functional categories of AD-related differentially expressed genes for two gene clusters, Cluster B and Cluster D. The analysis considers different functional categories, including Biological Process (BP), Cellular Component (CC), and Molecular Function (MF), for each cluster.}
    \label{fig091}
\end{figure*}

To evaluate the significance of the aforementioned signature gene classes in distinguishing early, middle, and late stages of AD, we conducted GO/KEGG enrichment analyses to examine their associations. Specifically, for the subsequent analysis, we primarily utilized the GO enrichment method provided by \citet{bu2021kobas}. Figure~\ref{fig091} illustrates the enrichment bubble map for GO enrichment analysis of two classes of AD-related differentially expressed gene sets, ClusterB and ClusterD. The enrichment bubble map primarily displays the top 10 functional categories in biological process (BP), cellular component (CC), and molecular function (MF) that are most abundant in the two gene classes. Additionally, Table~\ref{tab:5} presents the results of enrichment analysis for specific genes within these gene classes. Our analysis focuses on examining the characteristics of the functional categories in the two gene classes, as well as utilizing select genes to support our findings. More detailed results of the enrichment analysis can be found in the attached file (Enrichment\_analysis\_2datasets.xlsx).

The enrichment bubble plot of the ClusterB gene class (Figure~\ref{fig091}, Panel 1) reveals that its biological process (BP) functional categories primarily revolve around the development and synaptic transmission processes of the nervous system, as well as regulatory mechanisms associated with neurotransmitters and synaptic structure. In terms of cellular component (CC) function, the emphasis lies on extracellular structures and intracellular organelles and complexes, including collagen trimers, the extracellular matrix, extracellular space, as well as the nucleus and endoplasmic reticulum within cells. Regarding molecular function (MF), the focus is on the structural components of the extracellular matrix, protein binding, as well as functions related to enzymatic activity and cytokine activity. These functional categories align with the underlying pathological features of AD, specifically the ATN diagnostic framework \citep{jack2018nia,scheltens2021alzheimer}, suggesting the validity of the aforementioned enrichment analysis results.  


As previously mentioned, we posit that the principal components of the ClusterB differential gene class play a crucial role in discerning patients with early and middle-stage AD from those in the late-stage. In the late stages of AD, clinical manifestations become more pronounced, observable macroscopic alterations in brain biological traits occur, and microscopic protein detection can distinctly identify the advanced symptoms of AD \citep{hane2017recent}. Based on the results presented here, it is also possible to effectively distinguish patients with moderate-advanced AD by observing related traits controlled by genes in the ClusterB class. For example, \citet{misawa2008close} showed a potential association between gene AQP1 in astrocytes and A$\beta$ deposition in the AD brain, and senile plaques containing amyloid-beta peptide $A\beta_{1\text{-}42}$ are the major species present in the pathogenesis of AD. Similarly, regulation of genes COL4A1, FOS, and NAMPT had been linked to A$\beta$ deposition and potentially contributed to AD, with COL4A1 and FOS mainly acting through provoking inflammatory reactions \citep{marchesi2016gain,marchesi2011alzheimer,xu2021multimodal,xie2019nicotinamide}. \citet{moreno2020frontal} found that a variety of chitinase genes (including gene CHI3L2) were inflammatory biomarkers of AD.

The enrichment bubble plot of the ClusterD gene class (Figure~\ref{fig091}, Panel 2) reveals that its BP function involves multiple levels such as cell signal transduction and physiological regulation. The CC function focuses on neural cell structure \citep{otero2022molecular}, while the MF function involves different types of binding and regulatory activities. In contrast, the functional categories of the ClusterB gene class primarily focus on a certain aspect, whereas the functions of the ClusterD gene class cover a wider range.

Based on the analysis results of the decision tree, we posit that the principal component of the ClusterD differential gene class is a significant feature in distinguishing patients with early and middle-stage AD.
The results of ClusterD enrichment analysis (lower part of Table~\ref{tab:5}) revealed a key link between AD and the regulation of clathrin-dependent endocytosis and protein binding, that is, ClusterD class genes that can effectively distinguish between early and middle stage AD patients mainly control protein binding and neuronal synapse and neuronal cell-related metabolic processes. At this time, the clinical features of the patient are not obvious, but the internal metabolism of the brain has undergone microscopic changes \citep{porsteinsson2021diagnosis,dubois2016preclinical}. For example, as a neurodegenerative disease affecting cortical regions of the brain, abnormal presynaptic activity is also a potential feature affecting AD. Genes in the ClusterD class, namely ERC2 and SLC17A7, are associated with presynaptic cell function \citep{martinez2020canonical,sragovich2019autism}, while SNAP91 and SH3GL2 genes are involved in functions related to synaptic vesicles \citep{nguyen2019synaptic,hu2020co}. In addition, \citet{zhang2023bestrophin3} found that cerebral vascular smooth muscle cells were significantly reduced in AD patients, and the enrichment results of ClusterD gene class showed that MEF2C gene was involved in vascular associated smooth muscle cell migration, which verified the findings of this study \citep{maisuria2023conditional}. \citet{byman2019potential} found that abnormal glycogen catabolism can also affect neuronal cell metabolism, which represents another potential pathogenic factor in AD, and this result also corresponds to the PGM2L1 gene in Table~\ref{tab:5} \citep{morava2021impaired}.

Furthermore, through enrichment analysis of the genes, we observed a close association between oxygen carrier activity (IPCEF1) and protein heterodimerization activity (MEF2C and GABBR2) with both early and middle stages of AD. The formation of dimeric structures by AD-associated genes allows for synergistic interactions among mutated monomers, resulting in more functional enzyme forms that contribute to the maintenance of brain functionality. Protein heterodimerization activity associated with MEF2C and GABBR2 genes may disrupt this functional stability, ultimately affecting AD \citep{udeochu2023tau}. Oxygen carrier activity related to the gene IPCEF1 influences various aspects of normal cellular activities and represents a potential pathogenic factor in AD \citep{huang2019incorporating}.

\begin{table*}[h]
    \centering
    \vspace*{-6pt}
    \caption{ClusterB and ClusterD Gene enrichment analysis. Examples of some genes.}
\begin{tabular}{@{}ccccc@{}}
    \hline
\multicolumn{1}{l}{Class of genes} & GO term    & Representative genes                                      & Term                                                                    & Corrected P-Value \\
\hline
\multirow{6}{*}{ClusterB}          & GO:0038063 & COL4A1, COL4A2                                            & collagen-activated tyrosine kinase receptor signaling pathway           & 0.0049            \\
                                    & GO:0045944 & RORB, NAMPT, IL33, FGF1, FOS                              & positive regulation of transcription by RNA polymerase II               & 0.0263            \\
                                    & GO:0016807 & AQP1                                                      & cysteine-type carboxypeptidase activity                                 & 0.0421            \\
                                    & GO:0061304 & COL4A1                                                    & retinal blood vessel morphogenesis                                      & 0.0421            \\
                                    & GO:0004568 & CHI3L2                                                    & chitinase activity                                                      & 0.0421            \\
                                    & GO:0034356 & NAMPT                                                     & NAD biosynthesis via nicotinamide riboside salvage pathway              & 0.0440            \\
\hline
\multirow{11}{*}{ClusterD}         & GO:0005515 & \makecell{MEF2C, HPCAL1, VSNL1, ERC2,\\GABBR2, SNAP9, IPCEF1, SH3GL2 }   & protein binding                                                         & 0.0060            \\
                                    & GO:2000369 & SNAP91, SH3GL2                                            & regulation of clathrin-dependent endocytosis                            & 0.0060            \\
                                    & GO:0048488 & SNAP91, SNCB                                              & synaptic vesicle endocytosis                                            & 0.0155            \\
                                    & GO:0048786 & ERC2, SLC17A7                                             & presynaptic active zone                                                 & 0.0155            \\
                                    & GO:0030672 & SLC17A7, SH3GL2                                           & synaptic vesicle membrane                                               & 0.0307            \\
                                    & GO:0048790 & ERC2                                                      & maintenance of presynaptic active zone structure                        & 0.0340            \\
                                    & GO:1904753 & MEF2C                                                     & \makecell{negative regulation\\ of vascular associated smooth muscle cell migration} & 0.0340            \\
                                    & GO:0005344 & IPCEF1                                                    & oxygen carrier activity                                                 & 0.0350            \\
                                    & GO:0005980 & PGM2L1                                                    & glycogen catabolic process                                              & 0.0365            \\
                                    & GO:0045921 & VSNL1                                                     & positive regulation of exocytosis                                       & 0.0401            \\
                                    & GO:0046982 & MEF2C, GABBR2                                             & protein heterodimerization activity                                     & 0.0428     \\
                                    \hline      
                                \end{tabular}
                                \label{tab:5}%
                                \vskip12pt
\end{table*}

\subsection{Repeatability Analysis: Exploring Influential genes in human hrain development}

To further validate the reproducibility of the STW-MD method in our study, the statistical analysis processes in human hrain development were implemented  identically. For convenience of analysis, we followed the division proposed in \citet{kang2011spatio} to categorize the human brain development data into three stages: fetal development (10 PCW $\leq$ Age < 38 PCW, referred to as AGE0), postnatal development (0 Y $\leq$ Age < 20 Y, referred to as AGE1), and adulthood (Age $\geq$ 20 Y, referred to as AGE2), where PCW represents post-conceptional weeks and Y represents postnatal years. Notably, among the available samples, data from at least three brain regions were obtained for 37 samples.

A total of 13,479 differentially expressed genes were identified from a pool of 15,210 genes during the pre-weighted screening. This finding suggests that the expression levels of the majority of genes undergo significant changes throughout the three stages of brain development. Consequently, we selected the top 5\% of pairwise $\log_2 FC$ absolute values from each of the three periods as the weighted genes, resulting in a final set of 2,003 differentially expressed genes after combining them. 
In the multi-step decision tree process, we conducted screening on the weighted data, resulting in the identification of 515 differentially expressed genes. The aforementioned genes were divided into four groups. Similarly, based on the results of the decision tree (Figure S.9), gene classes ClusterA, ClusterC, and ClusterD were found to be associated with distinct stages of brain development. 


Figure S.10 illustrates a bubble plot representing the enrichment analysis results for these three classes of differentially expressed gene sets associated with brain development. Additionally, Table S.6 displays the outcomes of the enrichment analysis for selected genes within these three gene categories. More detailed results of the enrichment analysis can be found in the attached file (Enrichment\_analysis\_2datasets.xlsx). The enrichment bubble plot in Figure S.10 clearly demonstrates distinct functional differences among the three gene classes. Regarding BP function, ClusterA is primarily involved in cell signaling pathways and developmental processes. ClusterC is associated with gene expression regulation and cell maturation, exhibiting a broader range of functions. ClusterD predominantly encompasses various processes of cell differentiation.In terms of CC function, ClusterA is primarily associated with enveloped body-related structures such as coated nests and small bodies. ClusterC focuses on the nucleus and cytoplasm. ClusterD encompasses a wide range of intracellular and extracellular organelles, exhibiting both breadth and comprehensiveness. Regarding MF function, ClusterA primarily involves the binding of various proteins, exhibiting a single but wide-ranging function. ClusterC primarily involves the combination of RNA and protein, with a narrower scope but in-depth content. ClusterD encompasses binding activity and catalytic activity, with complex functional types. These enriched functions essentially reflect key features of brain development from fetal development to adulthood \citep{gulsuner2013spatial,edde2021functional,rietman2020using}. 

In summary, based on the results of the decision tree (Figure S.9), cluster diagram (Figure S.8(b)), and enrichment bubble plot (Figure S.10), we believe that ClusterA is strongly associated with AGE0, ClusterC is the gene class that distinguishes AGE0 from AGE1, and ClusterD is the key gene class for distinguishing AGE2 from the other two periods. ClusterA may be implicated in the healthy development and protection of the fetal brain (AGE0) \citep{peyvandi2023fetal,li2020epigenomic}. For example, \citet{jo2014versatile} demonstrated the essential role of SOX9 as a determinant of cell fate during embryonic development. Its expression facilitates the differentiation of cells from all three germ layers into various specialized tissues and organs. Genes in ClusterC might be implicated in the development of brain region function, where different regions of the brain gradually acquire specific functions, such as language, movement, perception, and so on \citep{vijayakumar2018puberty, mychasiuk2016epigenetic}. The study conducted by \citet{prieto2021mirnas} emphasized the significance of DICER1 as a crucial gene involved in miRNA biogenesis. This finding underscores its critical role in early brain patterning and gene regulation. Additionally, \citet{roa2022dicer} proposed that DICER1 plays a vital role in puberty. ClusterD emerges as the key gene class that distinguishes AGE2 from the other two periods. During AGE2, brain development reaches a relatively stable and mature stage, yet it can still be influenced by various physiological and environmental factors \citep{tooley2021environmental, huppi2010growth}. \citet{luo2020silencing} demonstrated that the down-regulation of AKR1B1 gene expression could impede the proliferation, invasion, and migration of glioma cells, ultimately promoting cell apoptosis. This finding suggests a potential role for AKR1B1 in regulating glioma cell behavior. More Details are illustrated in Supplementary Material S4.

\section{Conclusion}
\label{s:conclusion}


This paper focuses on the study of differential gene expression during dynamic and highly regulated brain development, including abnormal development. To effectively capture the spatial similarity and temporal dependence between different brain stages and regions, as well as address gene heterogeneity among various brain regions, this paper proposes a two-step modeling framework. The framework is based on spatio-temporal weighting and multi-step decision trees. This framework enables the analysis of differential gene expression in brain regions with high heterogeneity issues. The application of this model to two distinct datasets on brain development (the AD dataset and the brain development dataset) demonstrates a high consistency with existing studies, providing valuable insights into the process of brain development and abnormal development. 

The major innovation of this study is to propose a new weighting method that combines spatial (all brain regions) and temporal (different stages of brain development) dimensions to explore the potential factors affecting or causing abnormal brain development. As proposed above, the two-step modeling framework constructed in this paper has good adaptability, that is, more appropriate indicators or algorithms can be selected to replace according to the data needs. For example, in the brain region weighting algorithm, in addition to the fold change weight (FC value) utilized in the current study, other quantitative metrics such as the $t$-statistic and $p$-value could potentially be evaluated individually or in conjunction as alternative weighting schemes. The integrated multi-step decision tree algorithm proposed in this paper, the weighted brain gene expression data can also be well integrated with existing differential gene analysis methods, such as generalized linear model (GLM) and Bayesian method for differential based analysis \citep{wang2022guidelines}. In addition, the current integration algorithm of the multi-step decision tree only considers the simple concatenation of common algorithms, which has a notable effect. However, each step of the integration algorithm offers numerous options for selection and combination that have not yet been explored. For instance, other prevalent differential gene screening methods (DESeq2 \citep{love2014moderated} and edgeR \citep{robinson2010edger}), clustering algorithms (Seurat \citep{hao2021integrated}, WGCNA \citep{langfelder2008wgcna}) and decision tree models (Random Forest \citep{liu2022research}, XGBoost \citep{park2023development}) could be evaluated within the ensemble for a more comprehensive assessment.

This paper thoroughly investigates the two questions regarding spatiotemporal data of brain development raised in the Introduction. It primarily focuses on addressing the first question, taking into account the high heterogeneity observed among brain regions. Additionally, the second problem can be effectively solved through a straightforward backward derivation using the STW-MD method. Although the results of this paper can be well verified with the existing literature, the research process of this paper is relatively rough. For example, the PCA algorithm is easy to ignore the fine structure in comparison, so the interpretation of the biological significance of differentially expressed genes may be too simple. Biological processes are highly complex, with intricate connections between different genes. The study does not provide a macro perspective to interpret the results from a broader biologic perspective and may have overlooked other causal factors in Alzheimer's disease that may be implicit in the results.

\section*{Acknowledgements}

Text Text Text Text Text Text Text Text text text text text.
\vspace*{-12pt}

\section*{Funding}

This work has been supported by the... Text Text  Text Text.\vspace*{-12pt}

\bibliographystyle{natbib}
\bibliography{document2}

\end{document}